\documentclass[aps,twocolumn,showpacs,superscriptaddress]{revtex4-1}
\usepackage{amsmath,amssymb}
\usepackage{graphics,graphicx}
\usepackage{dcolumn,bm}
\usepackage{psfrag}
\newcommand{\cu}
{\affiliation{Department of Physics, University of Calcutta, 
92 Acharya Prafulla Chandra Road, Kolkata 700009, India.}}
\newcommand{\tifr}
{\affiliation{
Department of Theoretical Physics, Tata Institute of Fundamental Research, Homi Bhabha Road, Mumbai 400 005, India.}}

\begin{document}
 
\title{Opinion dynamics model with weighted influence: Exit probability and dynamics}

\author{Soham Biswas}%
\tifr
\author{Suman Sinha}%
\cu
\author{Parongama Sen}%
\cu

\begin{abstract}
We introduce a stochastic model of binary opinion   dynamics
in which the opinions are determined by the size of the neighbouring domains.
The exit probability here shows a step function behaviour 
indicating the existence of a separatrix distinguishing two different regions of basin of attraction.
This behaviour, 
  in one dimension,  is in contrast to other  well known  opinion dynamics
models where no such behaviour has been observed so far.
The coarsening study of the model also yields  novel exponent values.
A lower value of persistence exponent is obtained in the present model, which involves stochastic dynamics,  
when compared to that in a similar type of model  with  deterministic dynamics. 
This apparently counter-intuitive result is justified using further analysis. 
Based on these results it is concluded that the  proposed model belongs to a unique dynamical class.

\end{abstract}


\maketitle

\section{Introduction}
\label{sec1}

Nonequilibrium dynamics has been a topic of intensive research 
over the last few decades. The fact that models displaying identical equilibrium 
behaviour can be differentiated  on the basis of 
critical relaxation, coarsening and persistence behaviour,   has 
enhanced the interest in this field. Traditionally  
critical and off critical  dynamical behaviour  were studied for magnetic systems
with different dynamical rules and constraints \cite{Hoha,bray}. More recently,  physicists 
have been able to  construct dynamical models in problems  which are interdisciplinary 
in nature,  where one can investigate how such models can be classified into 
different dynamical  classes. 

One very popular interdisciplinary topic is opinion dynamics,
where a number of models \cite{stau,qvote,cast_sat,kszjsz} have been proposed and studied by physicists, 
many of which can also be regarded as spin models.
The  dynamical rules in all the well studied  opinion dynamics models with binary opinions 
($\pm 1$) in one dimension lead to the consensus state. The question 
 whether the intrinsic dynamics for ordering are equivalent has been asked in 
context of the generalised $q$ state voter model \cite{qvote}, generalised Glauber models \cite{cast_sat} and  Sznajd models \cite{kszjsz,slanina}. 
One of the quantities  which is computed to resolve this important question is 
the exit probability 
(EP) which denotes the probability $f_{up}(x)$ that one
ends up in a configuration  with all opinions equal to 1 starting from  $x$ fraction of opinions equal to 1.
EP  has been shown to be identical \cite{cast_sat} in the 
generalised Glauber model and the Sznajd model; the latter was  originally claimed to have a different dynamical scenario. 
In the  voter model (or Ising Glauber model) in
one dimension, $f_{up}(x)$ is simply equal to $x$, corresponding to the 
 conservation in the dynamics \cite{cast_review}. In the nonlinear voter model, long ranged Sznajd model and long ranged Glauber model,  
it is a nonlinear continuous function of $x$ \cite{cast_sat,slanina,lamb-redner2008}. In all these cases, the results are 
also independent of finite system sizes indicating there is no 
scaling behaviour. 
The claim that the exit probability for Sznajd model is a continuous function   
has been  however  questioned in \cite{galmart}. But analytical and numerical study of $q$ state nonlinear voter model (Sznajd model corresponds to nonlinear voter model
with $q= 2$) show that EP is a continuous function of $x$ \cite{przy}.
Interestingly,  the mean field result was shown to be exact  in the nonlinear $q$ voter model which includes the Sznajd model and independent of the range \cite{cast_sat,przy}. 
On the other hand, in  two dimensions or on networks, the exit probability shows a step 
function behaviour in many models which has been interpreted as a   phase transition. 
 $f_{up}(x)$ also shows finite size dependence in that case
\cite{staufferetal,croki}.
However, strictly speaking, one should interpret this as the existence of a separatrix 
between two different regions of   basin of attraction - where the attractors are the states 
with    all opinions equal to  $1$ or $-1$.

In this  paper, we have proposed a model  (the weighted influence model, WI model hereafter) which shows completely different behaviour  as the 
EP has  a step function behaviour even in one dimension.  
The result  also shows clear deviation from  mean field 
theory, although the latter provides a reasonable first order estimate.  
The model includes one parameter which  allows one  to obtain  a 
relevant ``phase diagram'' and also show the presence of universal behaviour. 

Apart from studying the exit probability, we also investigate the dynamical behaviour of the WI model
by studying the  density of domain walls $D(t)$ and persistence probability
 ${\cal {P}}(t)$ as functions of time $t$ starting from an initial disordered state.
The latter  is the probability that a spin has not flipped till time $t$ \cite{deribray}.
It is known that in conventional coarsening processes,    $D(t) \propto t^{-1/z}$
and ${\cal {P}}(t)$ shows a   scaling behaviour ${\cal {P}}(t) \propto t^{-\theta}$ in many systems \cite{bray}
where $z$ and $\theta$ are  the dynamic (growth) and persistence exponents respectively.
By calculating $z$ and $\theta$, the dynamical class
of the model can be identified and compared to  models 
for which these exponents are known, for example, for the zero temperature Ising model in one dimension, $z = 2$ and $\theta = 0.375$ are 
exactly known  results. 

The rest of the paper is arranged as follows: Sec. \ref{sec2} describes the model.
Sec. \ref{sec3} discusses the mean field theory in the context of the present work.
Numerical results obtained from extensive simulations are presented in Sec. \ref{sec4}.
Dynamical properties of the model are given in Sec. \ref{sec5} and finally in Sec. \ref{sec6},
concluding remarks are made.

\section{Description of the model}
\label{sec2}

The WI model is  a stochastic model 
with  opinions taking  values $\pm1$,
and there is a bias towards one type of opinion controlled by a relative weight factor. 
 It cannot be true that an initial majority always wins (in that case the same candidate/party will  go on winning  elections every time) \cite{elections}
and a huge majority of people may give up to an initial minority view \cite{minor,galam_minority}. 
The relative weight factor included in the model takes care of this idea.
This weight factor represents the
     relative strength that may arise due to monetary factors, local
    factors, muscle power, larger accountability, traditional, religious or
cultural  influence,
    recent incidents which has a great impact on the
    population etc.
The model is described in the following way. 
    Let there be  two groups of people with  opposing opinion  in the 
immediate spatial neighbourhood
    of an individual. Under the influence of these two groups,  she/he
    will be under pressure to follow one group. The pressure  is proportional
    to the size of the group. 
    Denoting the two neighbouring opposing group sizes  as  $S_{1}$ and $S_{-1}$ for opinion =  $\pm 1$  respectively, 
    an individual takes up opinion $1$ with probability
\begin{equation}
P_{1}=\frac{S_{1}}{S_{1} + \delta S_{-1}},
\label{prob1}
\end{equation}
    where $\delta$ is the relative influencing ability of the two groups and can vary from zero to $\infty$.
Probability to take opinion value $-1$ is $P_{-1}= 1-P_{1}.$
The model is considered in one dimension.
In case  an individual with opinion $1$ ($-1$) has both nearest neighbours with opinion $-1$ ($+1$), her opinion will change deterministically. 
The unweighted model corresponds to $\delta = 1$ \cite{foot}.

In this model, the dynamics is completely stochastic. A quasi-deterministic model in which the
neighbouring domain sizes determine the state of a spin had been proposed earlier (BS model
hereafter) \cite{pssb1}. 
The BS model takes into consideration the
sizes of neighbouring domains in the dynamics and the larger neighbouring 
domain
always dictates the opinion, irrespective of its size.
Only in the case when the neighbouring domains are of equal size (which occurs rarely) is the dynamical rule stochastic.
In the BS model, $z \simeq 1$ and $\theta \simeq 0.235$ (both the exponents
are different from the Ising model). 
We find some interesting  effects
of the nature of the stochasticity in the WI model, 
especially regarding the 
persistence behaviour, 
which is revealed  when  
compared to   the BS model and Ising model dynamical  results.

Regarding the binary opinion values as Ising spin states (up/down), the
absorbing states are the all up/all down states in the WI model. Probability
of attaining these consensus states however depends on the value of $\delta$
instead of being simply 1/2 (as in the Ising or voter  model) even when one starts from a completely random initial configuration ($x=1/2$). 
In the limit $\delta = 0$,  all spins will
 be up as $P_{1} = 1$ for any initial value of $x$ while for 
$\delta \to \infty$, the final state will be all down for any value of $x \neq 1$. 
Thus the threshold values  $x_c$ for which the final state will be all up is  zero for $\delta = 0$ and 1 for $\delta \to \infty$.
The exit probability is trivially a step function in these extreme limits.
The  question is what happens for other values of $\delta$, including $\delta = 1$. 

\section{Mean field theory}
\label{sec3}
The dynamics can be studied in terms of 
the motion of domain walls as only the 
spins adjacent to  domain walls can flip. In a Glauber like process, one considers the flipping of a random spin 
in time $\Delta t$ with the time unit being such that $\Delta t L =1$, where $L$ is the total number of spins.
Initially, there will be many domains of size one, but they
will quickly vanish as it is a deterministic process. Assuming no domain of size one remains 
in the system and using a mean field approximation, one can write down a microscopic equation for the (average) fraction of up spins  at time
$t+\Delta t$ given that there was a  fraction  $x$ at $t$.
It may be noted that in this approximation,   the fluctuations in the flipping probabilities $P_1$ and $P_{-1}$ can be ignored and they can be  taken to be  site-independent.
The equation for $x(t+\Delta t)$ is then given by 
\begin{eqnarray}
x(t+\Delta t) & =& r(t) [ (x(t)-1/L)P_{-1} + x(t)P_1] \nonumber \\
&& + r(t) [(x(t)+1/L)P_1 +  x(t)P_{-1}] \nonumber \\
&&+ (1-2r(t)) x(t).
\end{eqnarray}
Here $r(t)$ is the density  of domain walls, $r(t) \leq 1/2$ when domains have length at least 2. $P_1, P_{-1}$ are also in general time dependent.
The first two terms on the right hand side correspond to cases where the up and down spin at the
boundary are chosen for flipping respectively while the last term is for the case when a 
spin within a domain is selected ($x$ remains same  in the last  case obviously). 
Thus one gets
\begin{equation}
\frac{dx}{dt} = r(t) [P_{1} - P_{-1}].
\label{fixed}
\end{equation}
This equation cannot be solved without knowing the dynamical equation for $r(t)$ which is again expected to   involve $x(t)$ in 
a complicated manner.  
However, the  fixed points of the 
equation, in which we are actually interested,  
  are easily obtained;
a trivial fixed point $r(t) = 0$ and the other one is $[P_{1} - P_{-1}] =0$.
For the Ising or voter model, $P_{1}$ is 
equal to $P_{-1}$ which corresponds to the  result that 
$dx/dt =0$ independent of $x$. This  leads to the known result that the exit probability is simply equal to $x$.  All points are fixed points here. 
  In case 
one gets a single fixed point $x_c$ from   Eq. (\ref{fixed}), it will indicate the existence of the  step function like behaviour  
associated with the exit probability.
  The mean field approximation    of course neglects all correlations and fluctuations.
In the WI model, $P_{-1}$ and $P_{1}$ in 
mean field approximation can be estimated by taking  $S_1$ and $S_{-1}$ 
proportional to $x$ and $(1-x)$ respectively 
(at the fixed point) in  Eq. (\ref{prob1}). 
There is no reason to take the  constant of proportionality to be  different (i.e. there is no bias to 
either type of  domain) 
such that  $P_1 = x/[x+\delta(1-x)]$ and we get
\begin{equation}
x_c= \delta/(1+\delta).
\label{eq-mf}
\end{equation}

 Although the mean field result  involves many assumptions it is tempting to accept this result as it coincides with the limiting  results 
that $x_c = 0$ for $\delta =0$, $x_c = 1.0$ for $\delta \to \infty$.
The mean field result also predicts $x_c = 1/2$ for $\delta = 1$. 
$\delta=1$  corresponds to the model with  unweighted influence, and here 
if one starts with $x=1/2$ the system will go to $+1$ state with $50\%$ probability (by the argument of symmetry). If there is any initial bias ($x \neq 1/2$) in
the system then it will win at the end. EP will be zero for $x<1/2$ and equal to 1 for  $x> 1/2$. Hence one expects  that at $\delta=1$, $x_c =1/2$ as 
given by (\ref{eq-mf}).

Having obtained the evidence of a single value of $x_c$  from mean field approximation, 
our next job is to find out numerically whether there exists a 
separatrix and  
and whether finite size effects exist. 
Also, the deviation from mean field theory, if any, will be investigated in the following section.

\section{Exit probability: numerical results}
\label{sec4}
 
\begin{figure}[!h]
\begin{center}
\resizebox{95mm}{!}{\includegraphics {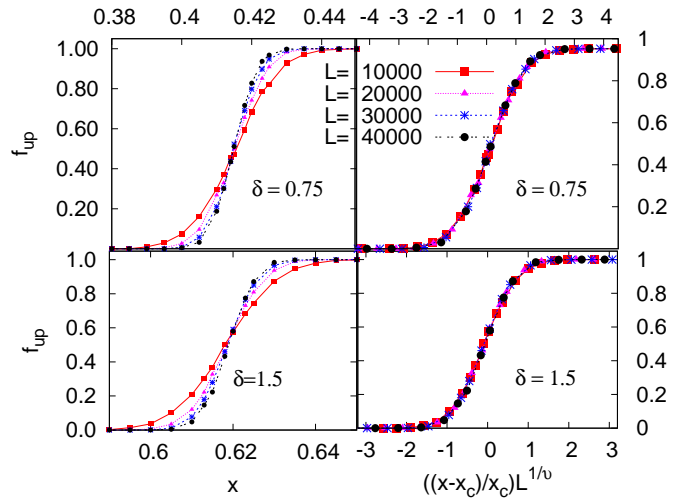}}
\end{center}
\caption{(Color online) The data for the exit probability against initial concentration $x$ (left panel) and the data collapse (right panel) 
 using $\nu=2.50 \pm 0.02$ (Eq. (\ref{epscaling}))
for different system sizes are shown. }
\label{epvsups}
\end{figure}

We calculate the exit probability  $f_{up}$
for system sizes ranging from $L=5000$ to $L=50000$ and repeat the simulations for over at least $3000$ configurations for 
each system size. 
$f_{up}$  against initial concentration $x$ is shown in
 Fig. \ref{epvsups} for two values of $\delta$. It indeed shows a sharp rise close to a   value of $x \simeq x_c$, henceforth
called the separatrix  point. 
The shape of the exit probability  $f_{up}$   plotted against $x$ immediately shows that it is nonlinear,  
moreover,  
  $f_{up}(x)$ curves show strong system size dependence and intersect at a single point $x_c$ for different values of $L$. 
The behaviour of EP indicates that 
it shows a step function behaviour in the thermodynamic limit.
Finite size scaling analysis can  be made using  the   scaling form (when $\delta$ is constant):
\begin{equation}
f_{up}(x,L)=f_1\left(  \frac{(x-x_c)}{x_c}{L^{1/\nu}}\right) 
\label{epscaling}
\end{equation}
where  $f_1(y) \to 0$  for $y << 0$  and 
equal to 1 for $y >> 0$  (i.e.,   a step function in the thermodynamic limit).  
$f^c_{up}$ gives the the value of EP at the separatrix point.   
The data collapse, shown in Fig. \ref{epvsups}, takes place   
with  $\nu=2.50 \pm 0.02$ for all values of $\delta$. 
For fixed $x$, the finite size scaling form for exit probability can be written as:
\begin{equation}
f_{up}(\delta,L)=f_2 \left( \frac  {(\delta -\delta_c)}{\delta_c}{L^{1/\nu}} \right).
\label{delscaling}
\end{equation}
where $f_2(y) \to 0$  for $y >> 0$, equal to 1 for $y << 0$.
Both the scaling forms (Eqs (\ref{epscaling}) and (\ref{delscaling})) give $\nu =2.50 \pm 0.02$ 
independent of the exact location of the separatrix  point. 
$x_c$ as a function of $\delta$ denotes the trajectory of the separatrix  point 
as $\delta$ is varied and for convenience we call it the   ``phase boundary''.
So one can conclude that  universal behaviour exists along the entire phase boundary.

\begin{figure}[!h]
\begin{center}
\resizebox{85mm}{!}{\includegraphics[scale=1.2]{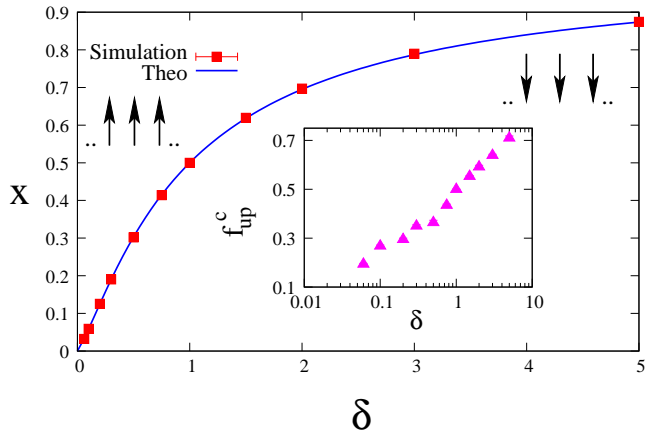}}
\end{center}
\caption{(Color online) Main plot:
The phase boundary in the $x-\delta$ plane obtained from simulation and its theoretical fitting (Eq. (\ref{eqcorec})) separating the all up and all down phases.
Inset shows  variation of $f^c_{up}$ with  $\delta$. } 
\label{delvsep}
\end{figure}

Estimating $x_c$ for  different values of $\delta$, we plot the phase boundary in 
 Fig. \ref{delvsep}.
\begin{figure}[!h]
\begin{center}
\resizebox{85mm}{!}{\includegraphics[scale=1.0, angle=270]{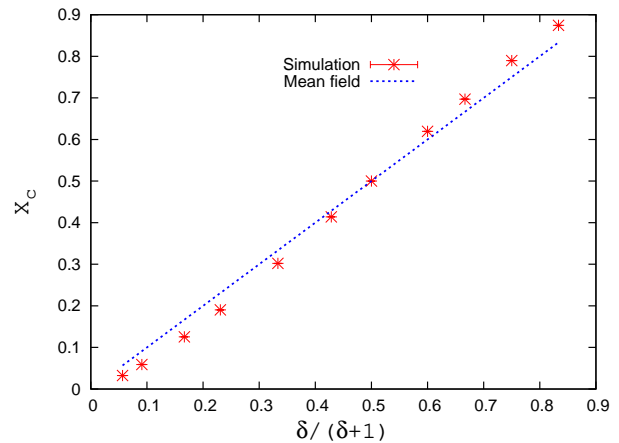}}
\end{center}
\caption{(Color online) 
The values of $x_c$ obtained by simulations and the mean field theory result (dashed line, Eq. (\ref{eq-mf}) are compared. There is appreciable difference 
 away from $\delta =1$; the difference vanishes as $\delta  \to 0$ and $\delta \to \infty$.  }
\label{meanfield}
\end{figure}
The phase boundary is not exactly given by  the mean field estimate  
(\ref{eq-mf})  but shows systematic  deviation from this equation  (except at  $\delta = 0, 1.0$ and $\delta \to \infty$) as shown in Fig. \ref{meanfield}.
There is a systematic difference  from the mean field results away from $\delta =1$ which vanishes at $\delta =0 $ and also as $\delta \to \infty$.
However, the difference is mostly less than ten percent for  other 
values of $\delta$ and
therefore the mean field result can be taken as a first order estimate.
In principle,  one may consider an expression of $x_c$ given by a polynomial in
$\delta/(1+\delta)$  as obviously it deviates from a simple linear form. 
However, introducing only a second order  term is not sufficient and 
we therefore attempt to 
fit the numerically obtained values of $x_c$ accurately by a single correction
term, the form of which is conjectured by the known values of $x_c$ at 
$\delta = 0,1 $ and $\delta \to \infty$.
  We  assume   the following form for $x_c$: 
\begin{equation}
x_c = \delta/(\delta+1) + a \delta(\delta-1)/(\delta +b)^c.
\label{eqcorec}
\end{equation}
Here $\delta(\delta-1)$ in the correction term (second term on right hand side 
of Eq. (\ref{eqcorec})) takes care that the term vanishes for
$\delta = 0,1$. If one compares the numerical data with the mean field result (Fig. \ref{meanfield}), 
the former  gives larger values of   $x_c$  for $\delta> 1$ (and lower values of $x_c$ for $\delta < 1$).
So $(\delta-1)$ will be there in the correction term instead of $(1-\delta)$.  
One also needs a factor proportional to a power of $\delta$ in the denominator
of the correction term which should be nonzero for $\delta = 0$ and make the term vanish in the 
limit $\delta \to \infty$. Such a term is chosen as $(\delta + b)^c$ 
where $c$ should be greater than $2$ and $b\neq 0$.
We indeed find that Eq. (\ref{eqcorec})  fits the curve quite nicely with $a=0.18 \pm 0.02, b=0.67 \pm 0.04$ and $c= 2.59 \pm 0.06$.

We also investigate the behaviour of $f^c_{up}$ $\equiv$ $f_{up}(x_c)$ as a 
function of $\delta$; although a monotonic increase is 
found, no obvious functional  form   appears to fit the data.

\section{Dynamical properties}
\label{sec5}
Next we consider the dynamical behaviour 
by studying the  density of domain walls $D(t)$ and persistence probability  
 ${\cal {P}}(t)$ 
 as functions of time $t$.
In this context, comparison with the zero temperature dynamics in  the 
Ising model and the BS model \cite{pssb1}  will be interesting.

Coarsening  study for    $x=0.5$: 
We start with a completely disordered state where $\delta = 1$ is the 
estimated separatrix  point. 
The scaling behaviour of  $D(t)$  is compatible with a value of   $z =1$ at  $\delta=1$.   As $\delta$ deviates from $1$,
the coarsening process becomes very fast: obviously, in the extreme limit $\delta = 0$ or $\infty$, 
the system goes to the all up/down configuration almost instantaneously. In fact for any value of $\delta \neq 1$, the power law behaviour for $D(t)$ is no longer valid. 
This is not surprising, it is known that power law scalings are valid only on the transition point (e.g., in the Ising model,
the order parameter shows exponential decay to its equilibrium value away from the critical temperature).

\begin{figure}[!ht]
\begin{center}
\resizebox{80mm}{!}{\includegraphics[scale=1.7]{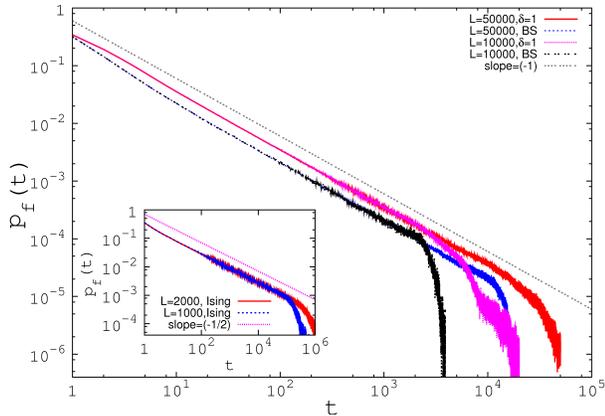}}
\end{center}
\caption{(Color online) Decay of average flipping probability $P_f(t)$ with time for both the WI model (with $\delta=1$) and the BS model with initial fraction of up spins $x = 1/2$.
The results are shown for system sizes $L = 10000$ and 50000. Finite
size effects  appear only at very large times where an exponential cutoff 
appears 
due to the finite size.  
Slope of the dashed line is $-1$.
Inset shows the decay of $P_f(t)$   with time for the nearest neighbour Ising model. Slope of the dashed line is $-0.5$.}
\label{flip}
\end{figure}

The growth exponent $z$ in  coarsening phenomena in spin systems 
can be found out by studying
several quantities apart from the domain density $D(t)$ which varies as $t^{-1/z}$. These include the variations of the absolute magnetisation
with time, total time
to reach equilibrium as a function of system size and the fraction of spin flips
  again as a function of time $t$.
The last quantity,
 $P_f(t)$,   is expected to follow the
same behaviour as $D(t)$ since spins at the domain boundary can flip
only (it will be less than $D(t)$ in magnitude though). 
This is true for all models.
For  $x = 1/2$ and $\delta = 1$,   in the WI model $P_f(t)$
thus  varies with time as $\sim t^{-1/z}$ with $z \simeq 1$.
We have shown in Fig. \ref{flip} the scaling behaviour of the flipping probability
as this quantity is useful in understanding the persistence behaviour. 
  
\begin{figure}[!ht]
\begin{center}
\resizebox{85mm}{!}{\includegraphics[scale=2.3]{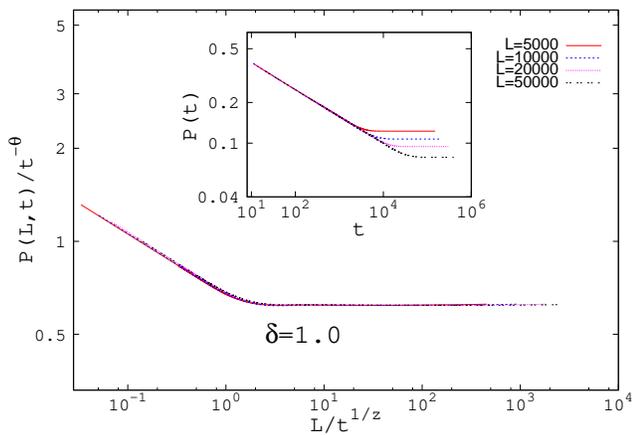}}
\end{center}
\caption{(Color online) The collapse of the scaled persistence probability against scaled time using
$\theta=0.20 \pm 0.002$ and $z=1.0 \pm 0.002$ for different system sizes. Initial fraction of up spin $x=1/2$. Inset shows the unscaled data.}
\label{colper}
\end{figure}

The following dynamical scaling form for the persistence probability is  used \cite{prgm,annni} to obtain both $z$ and $\theta$.
\begin{equation}
{\cal{P}}(t,L) \propto t^{- \theta}f(L/t^{1/z}).
\label{persca}
\end{equation}
The persistence probability saturates at a value $\propto L^{\alpha}$ at large times in finite systems where $\alpha$ is related to the spatial 
correlation of the persistent spins at $t \to \infty$.
So the scaling function $f(y) \sim y^{-\alpha}$ with $\alpha =z\theta$ for $y << 1$ and $f(y)\rightarrow$ constant at large $y$.
We have estimated the exponents $\theta$ and  $z$ from the above scaling relation for $\delta =1$,
giving  $\theta=0.20 \pm 0.002$  and $z=1.0 \pm 0.002$. The raw data as well as the scaled data are  shown in Fig. \ref{colper}.
These exponents are universal in the sense that if one starts at any point on the phase boundary (i.e., with the initial fraction of up spins equal to $x_c(\delta)$), 
one gets the same values.  

In  the WI model,   $z \simeq 1$  as  in the  BS model but the 
persistence exponent is different (by more than ten percent numerically). In fact, $\theta \sim 0.20$ obtained here is  
less than the BS model value ($\sim 0.235$), which is a bit counter-intuitive as the WI  model is completely stochastic while the BS model
is not. 
In the BS model, there is a ballistic motion of the surviving domain walls,
which in the course of their motion will flip all the spins that appear on one of their boundaries. With high probability,
these spins will flip once only  so that if any flipping occurs it is more likely to  affect the persistence probability. 
 In the WI  model, there will be motion of the 
domain walls in both directions (though  less in comparison   to the Ising model where pure random walk is executed before the 
domain walls are annihilated), 
such that the same spin may  flip more than once with 
higher probability and obviously persistence
will not be affected when a spin flips more than once. To check this, we have computed the 
distribution $g(n)$ of the number ($n$) of times a spin flips in both the models. 
\begin{figure}[!ht]
\begin{center}
\resizebox{80mm}{!}{\includegraphics[scale=1.2]{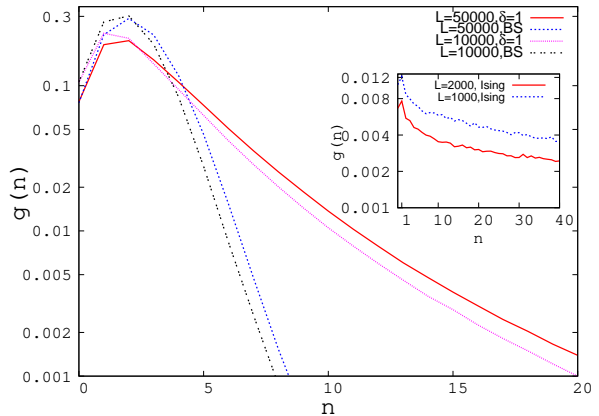}}
\end{center}
\caption{(Color online) The probability $g(n)$  that a spin flips  $n$ number of times up to the time of reaching the equilibrium state is plotted against
$n$ for both the BS model and the WI model (with $\delta=1$). Inset shows the same for the nearest neighbour
Ising model.}
\label{probdistri}
\end{figure}
The results, shown in Fig. \ref {probdistri}, exhibit indeed that in the BS model, 
$g(1)$ is much higher compared to the WI model
while probability that a spin flips   
a large number of times ($n > 4$) is much less. (Roughly, for $n>3$, $\log g(n)  \propto {- n}$ for the BS model and $\propto  - \sqrt n$
for the WI  model with $\delta =1$.)
We have already seen that the probability of flipping $P_f(t)$  
shows the same scaling behaviour in the two models and
this shows why the  persistence decays in a slower manner in the  present model. 

 In comparison, in the Ising model, the persistence probability 
decays fastest although the domain walls perform pure random walk. 
Here, $P_f(t) \propto t^{-1/2}$ implies that   domains  survive much longer 
and  as a  result a larger number of  spins are flipping at every step. 
 So although $g(n)$ has a much slower decay 
the persistence probability decays  much faster 
showing that the effect of the slower decay in  the number of domain walls  
 is more important than the unbiased random walk motion of each. 

Stochasticity in the BS model has been introduced in several ways  earlier by incorporating a parameter in the 
model \cite{sbpspr,ps1}. In fact in \cite{ps1}, the domain sizes had been considered in the dynamics in a different manner. 
 However, even such stochasticity in the BS model could not lead to any new dynamical
behaviour ($z$ and $\theta$ were found to be equal to $\sim 1$ and $\sim 0.235$ respectively).
On the other hand, the stochastic model considered here shows a different result for $\theta$ even for  the unweighted model ($\delta = 1$).

\section{Concluding remarks}
\label{sec6}
We have presented a stochastic binary opinion dynamics model with one parameter where 
the  exit probability has a step function like behaviour  even in  
one dimension, in contrast to other familiar models.  
One obtains a separatrix which is   similar to that appearing in magnetic 
systems at zero field, separating regions of positive and negative 
magnetisation although in the latter  one considers strictly the  
equilibrium behaviour.
The results show finite size scaling where the scaling argument 
is $|x-x_c|L^{1/\nu}$ indicating that the width of the region where $f_{up}$
is not equal to unity 
  or zero decreases as $L^{-1/\nu}$.  

The unique behaviour of the exit probability    may  be present
 due to the 
  effective long range interactions in the WI model.  
 However it has been shown previously for the generalised voter and Sznajd models that
 the exit probability does not change its nature even if one  makes the range of interaction infinite \cite{cast_sat,przy}. This indicates that the dynamical rule of WI model which handles the range of interaction in a subtly different manner 
could be responsible for  the behaviour of the exit probability.  
A thorough study of similar models with domain size 
dependent dynamics is in progress to check this   \cite{soham}.
One may also attempt to check the dependence of $\nu$ when the problem
is considered in higher dimensions.
It is also observed that there is a deviation from the mean field result 
unlike other models in one dimension. 
This deviation is attributed to the fact that the fluctuations which
have been ignored (e.g., by taking $P_1, P_{-1}$ independent of location and 
replacing all  domain sizes by an average value) are indeed relevant.
 However, the deviation from mean field estimates is still small such that 
the mean field result can be considered as a first order calculation.

 The other important and interesting result
is that the persistence exponent of WI model 
is not only different but lowest    among the well known 
models, including those where domain size dependent dynamics 
have been used.  Thus the  WI model 
is claimed to  belong to a unique dynamical class in opinion dynamics models. 

\section{Acknowledgement}

We are grateful to Deepak Dhar for a critical reading of an earlier version of the manuscript and for some useful discussions.
SS acknowledges support from the UGC Dr. D. S. Kothari Postdoctoral Fellowship under grant No. F.4-2/2006(BSR)/13-416/2011(BSR). 
SB thanks the Department of Theoretical Physics, TIFR, for the use of its computational resources.
PS thanks UPE (UGC) project for computational  support.

\end{document}